\documentstyle[preprint,seceq]{jpsj}

\def\runtitle{Backward Scattering and Coexistent State 
in Two-Dimensional Electron System}
\def\runauthor{Masakazu {\sc Murakami} and Hidetoshi {\sc Fukuyama}}

\title
{Backward Scattering and Coexistent State\\ 
in Two-Dimensional Electron System}
\author
{ 
Masakazu {\sc Murakami}\footnote{E-mail: murakami@watson.phys.s.u-tokyo.ac.jp}
and Hidetoshi {\sc Fukuyama}
}

\inst
{Department of Physics, Tokyo University, Tokyo 113}

\recdate
{February 2, 1998}

\abst
{The results of the mean field studies on the effects of 
the backward scattering 
with large momentum transfer 
in a two-dimensional electron system 
are extended to the case with various types of the Fermi surface
and coupling constants.
It is found that the coexistent state of $d$-wave superconductivity,  
antiferromagnetism and $\pi$-triplet pair can be stabilized
quite generally near half filling.}

\kword
{two-dimensional system, saddle-point singurality, backward
scattering, $d$-wave superconductivity, antiferromagnetism, $\pi$-triplet pair
}

\begin{document}
\sloppy
\maketitle
\newread\epsffilein    
\newif\ifepsffileok    
\newif\ifepsfbbfound   
\newif\ifepsfverbose   
\newif\ifepsfdraft     
\newdimen\epsfxsize    
\newdimen\epsfysize    
\newdimen\epsftsize    
\newdimen\epsfrsize    
\newdimen\epsftmp      
\newdimen\pspoints     
\pspoints=1bp          
\epsfxsize=0pt         
\epsfysize=0pt         
\def\epsfbox#1{\global\def\epsfllx{72}\global\def\epsflly{72}%
   \global\def\epsfurx{540}\global\def\epsfury{720}%
   \def\lbracket{[}\def\testit{#1}\ifx\testit\lbracket
   \let\next=\epsfgetlitbb\else\let\next=\epsfnormal\fi\next{#1}}%
\def\epsfgetlitbb#1#2 #3 #4 #5]#6{\epsfgrab #2 #3 #4 #5 .\\%
   \epsfsetgraph{#6}}%
\def\epsfnormal#1{\epsfgetbb{#1}\epsfsetgraph{#1}}%
\def\epsfgetbb#1{%
%
%
\openin\epsffilein=#1
\ifeof\epsffilein\errmessage{I couldn't open #1, will ignore it}\else
%
%
   {\epsffileoktrue \chardef\other=12
    \def\do##1{\catcode`##1=\other}\dospecials \catcode`\ =10
    \loop
       \read\epsffilein to \epsffileline
       \ifeof\epsffilein\epsffileokfalse\else
%
%
          \expandafter\epsfaux\epsffileline:. \\%
       \fi
   \ifepsffileok\repeat
   \ifepsfbbfound\else
    \ifepsfverbose\message{No bounding box comment in #1; using defaults}\fi\fi
   }\closein\epsffilein\fi}%
%
%
\def\epsfclipon{\def\epsfclipstring{ clip}}%
\def\epsfclipoff{\def\epsfclipstring{\ifepsfdraft\space clip\fi}}%
\epsfclipoff
\def\epsfsetgraph#1{%
   \epsfrsize=\epsfury\pspoints
   \advance\epsfrsize by-\epsflly\pspoints
   \epsftsize=\epsfurx\pspoints
   \advance\epsftsize by-\epsfllx\pspoints
%
%
   \epsfxsize\epsfsize\epsftsize\epsfrsize
   \ifnum\epsfxsize=0 \ifnum\epsfysize=0
      \epsfxsize=\epsftsize \epsfysize=\epsfrsize
      \epsfrsize=0pt
%
%
     \else\epsftmp=\epsftsize \divide\epsftmp\epsfrsize
       \epsfxsize=\epsfysize \multiply\epsfxsize\epsftmp
       \multiply\epsftmp\epsfrsize \advance\epsftsize-\epsftmp
       \epsftmp=\epsfysize
       \loop \advance\epsftsize\epsftsize \divide\epsftmp 2
       \ifnum\epsftmp>0
          \ifnum\epsftsize<\epsfrsize\else
             \advance\epsftsize-\epsfrsize \advance\epsfxsize\epsftmp \fi
       \repeat
       \epsfrsize=0pt
     \fi
   \else \ifnum\epsfysize=0
     \epsftmp=\epsfrsize \divide\epsftmp\epsftsize
     \epsfysize=\epsfxsize \multiply\epsfysize\epsftmp   
     \multiply\epsftmp\epsftsize \advance\epsfrsize-\epsftmp
     \epsftmp=\epsfxsize
     \loop \advance\epsfrsize\epsfrsize \divide\epsftmp 2
     \ifnum\epsftmp>0
        \ifnum\epsfrsize<\epsftsize\else
           \advance\epsfrsize-\epsftsize \advance\epsfysize\epsftmp \fi
     \repeat
     \epsfrsize=0pt
    \else
     \epsfrsize=\epsfysize
    \fi
   \fi
%
%
   \ifepsfverbose\message{#1: width=\the\epsfxsize, height=\the\epsfysize}\fi
   \epsftmp=10\epsfxsize \divide\epsftmp\pspoints
   \vbox to\epsfysize{\vfil\hbox to\epsfxsize{%
      \ifnum\epsfrsize=0\relax
        \includegraphics{\ifepsfdraft}%
      \else
        \epsfrsize=10\epsfysize \divide\epsfrsize\pspoints
        \includegraphics{\ifepsfdraft}%
      \fi
      \hfil}}%
\global\epsfxsize=0pt\global\epsfysize=0pt}%
%
%
{\catcode`\%=12 \global\let\epsfpercent=
%
%
\long\def\epsfaux#1#2:#3\\{\ifx#1\epsfpercent
   \def\testit{#2}\ifx\testit\epsfbblit
      \epsfgrab #3 . . . \\%
      \epsffileokfalse
      \global\epsfbbfoundtrue
   \fi\else\ifx#1\par\else\epsffileokfalse\fi\fi}%
%
%
\def\epsfempty{}%
\def\epsfgrab #1 #2 #3 #4 #5\\{%
\global\def\epsfllx{#1}\ifx\epsfllx\epsfempty
      \epsfgrab #2 #3 #4 #5 .\\\else
   \global\def\epsflly{#2}%
   \global\def\epsfurx{#3}\global\def\epsfury{#4}\fi}%
%
%
\def\epsfsize#1#2{\epsfxsize}
%
%
\let\epsffile=\epsfbox

\section{Introduction}
In recent years the electronic states in two-dimensional systems 
in the copper oxide high-$T_{\rm c}$ superconductors 
have been studied intensively.
Especially in the normal state near the optimal doping, 
a very flat dispersion of quasiparticle excitations 
around ($\pi$,0) and (0,$\pi$),
i.e., the extended saddle-point singurality, 
has been revealed in the angle resolved photoemission spectroscopy (ARPES)
experiments.\cite{ARPES1,ARPES2}
This flatness of dispersion
is more prominent than that obtained in the single-particle
band calculations such as local-density approximation (LDA).\cite{LDA1,LDA2}
This behavior has been attributed to the many-body
correlation effects based on the results 
of quantum Monte Calro (QMC) simulations,
\cite{Bulut}
and propagator-renormalized fluctuation-exchange (FLEX) approximation,
\cite{ca}
or to the dimple of the CuO$_{2}$ planes from the recent 
LDA calculations.\cite{LDA3}
It is the well-known fact that saddle points in the electronic structure 
result in the van Hove singuralities (vHs) 
in the electronic density of states (DOS).
Especially for two-dimensional case, they produce logarithmic divergence
in the DOS. Therefore, their existence near the Fermi level
has important physical consequences.
 
In the hole-doped case, such as YBa$_{2}$Cu$_{3}$O$_{7-\delta}$ (YBCO) 
or Bi$_{2}$Sr$_{2}$CaCu$_{2}$O$_{8+\delta}$ (BSCCO), 
the Fermi level lies near ($\pi$,0) and (0,$\pi$).\cite{ARPES2}
This might be responsible for unconventional physical properties,
such as the normal-state resistivity with $T$-linear 
temperature dependence\cite{resistivity} or 
the superconducting gap of the $d_{x^{2}-y^{2}}$ symmetry.
\cite{aniso,junc}
On the other hand, in the underdoped region, 
a pseudogap of the $d_{x^{2}-y^{2}}$ symmetry,
which is consistent with the superconducting gap symmetry, 
has been observed around ($\pi$,0) and (0,$\pi$) 
above the superconducting critical temperature.\cite{pgap1,pgap2,pgap3}
It has been indicated that there is strong coupling between 
quasiparticle excitations near the flat band and 
collective excitations centered near ($\pi$,$\pi$).\cite{pgap4}

In the electron-doped case, 
such as Nd$_{2-x}$Ce$_{x}$CuO$_{4+\delta}$ (NCCO), 
however, the Fermi level lies 
far above ($\pi$,0)
and (0,$\pi$), and instead gets close to ($\pm \pi /2$, $\pm \pi /2$).
\cite{ARPES2,ncco1,ncco2}
The Fermi surface agrees very well with LDA results.\cite{LDAncco}
This might be related to conventional (or 
Fermi-liquid type)
physical properties, such as normal-state resistivity with $T$-square
temperature dependence\cite{inplane} or the superconducting gap
which is {\em nodeless} on the Fermi surface 
(or BCS-like gap function).\cite{depth} 

Theoretically, 
various shapes of the Fermi surface
can be reproduced by introducing 
not only nearest-neighbor hopping $t$ 
but also next-nearest-neighbor hopping $t'$.
Studies looking for the instabilities
have been carried out 
with a special emphasis on ($\pi$,0) and (0,$\pi$)
in the 2D Hubbard model, 
by use of the renormalization group method
\cite{Schulz,ttu} 
and QMC.\cite{Husslein} 
These studies have shown that 
$d$-wave superconductivity prevails 
over antiferromagnetism by the effect of $t'$.
In our preceding letter,\cite{ore2}
we have studied possible ordered states
with a special emphasis on the effects of the {\em backward} scattering
processes with large momentum transfer  
between two electrons near ($\pi$,0) and (0,$\pi$) by
introducing $g_{1}$ and $g_{3}$ processes,
which correspond to 'exchange' and 'Umklapp' processes, 
respectively. The mean field phase diagram
for a special case, $g_{1}=g_{3}$, and for the YBCO type Fermi surface,
was determined.
In this paper, results of more detailed studies for general cases with
various shapes of the Fermi surface and for several choices of 
$g_{1}$ and $g_{3}$ will be presented.
In \S 2, we introduce the model Hamiltonian and three types of 
order parameters, i.e., $d$-wave Cooper pair,
N\'eel order and $\pi$-triplet pair.
In \S 3, we determine the phase diagram in the plane of temperature, $T$, 
and the hole (or electron) doping rate, $\delta$ (or $x$).
\S 4 is devoted to conclusion and discussion. 

We take unit of $\hbar=k_{\rm B}=1$.

\section{Model}
We consider a two-dimensional square lattice with the kinetic energy
given by,
\begin{eqnarray}
H_{0}&=& -\sum_{<ij>\sigma}t_{ij}\{c^{\dagger}_{i\sigma}c_{j\sigma}
+(\mbox{h.c.})\} - \mu\sum_{i}c^{\dagger}_{i\sigma}
c_{i\sigma},
\nonumber \\
&=& \sum_{p\sigma}\xi_{p}c^{\dagger}_{p\sigma}c_{p\sigma},
\label{eqn:h0}
\end{eqnarray}
where $t_{ij}$ is the transfer integral, $c_{i\sigma}(c^{\dagger}_{i\sigma})$
is the annihilation (creation) operator for the electron on the i-th site
with spin $\sigma$, and $\mu$ is the chemical potential.
The energy dispersion $\xi_{p}=\epsilon_{p}-\mu$ is given by
\begin{eqnarray}
\epsilon_{p}&=&-2t(\cos p_{x}+\cos p_{y})-4t'\cos p_{x}\cos p_{y}\nonumber\\
&&-2t''(\cos 2p_{x}+\cos 2p_{y}),
\end{eqnarray}
including $t$ (nearest neighbor), $t'$ (next nearest neighbor) and $t''$ 
(third neighbor), as shown in Fig.~\ref{hopping}.
We take $t$ as the energy unit, i.e., $t=1$ and the lattice constant is also 
taken as unity.
The energy band
$\epsilon_{p}$ has saddle points at ($\pm\pi$,0)
and (0,$\pm\pi$). 
The number of independent saddle points in the 1st Brillouin zone 
is equal to $2$.

\begin{figure}
\begin{center}
\leavevmode\epsfysize=3cm
\epsfbox{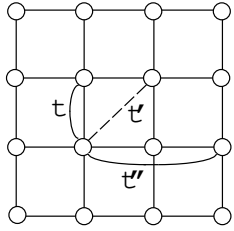}
\end{center}
\caption{The transfer integrals on a 2D square lattice;
$t$ (nearest neighbor), $t'$ (next nearest neighbor) and $t''$ 
(third neighbor).}
\label{hopping}
\end{figure}

The 1st Brillouin zone is usually taken
as shown in Fig.~\ref{2dbz} (a). In this case, however,
the saddle points lie on the zone boundaries. 
In order to treat the scattering processes between two electrons
around the saddle points unambiguously,
it is convenient to choose the 1st Brillouin zone 
inside which the saddle points lie,
as shown in Fig.~\ref{2dbz} (b). It consists of the region A and B 
including $Q_{A}\equiv(\pi,0)$ and $Q_{B}\equiv(0,\pi)$, respectively.
We note that $w_{p}\equiv \mbox{sgn}(\cos p_{y}-\cos p_{x})=+1(-1)$
for $p\in$A(B), which determines the symmetry of the superconducting order
parameter, as we shall see later.

\begin{figure}
\begin{center}
\leavevmode\epsfysize=5cm
\epsfbox{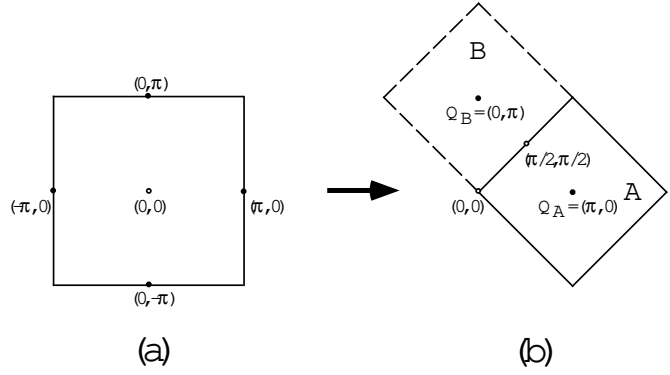}
\end{center}
\caption{The 1st Brillouin zone. (a) original one. (b)
our choice.}
\label{2dbz}
\end{figure}

Now we introduce effective interaction between two electrons 
in the region A and B, just as in the g-ology 
in one-dimensional electron system.
Here we treat only the backward scattering with large momentum 
transfer, shown in Fig.~\ref{g1g3}, i.e., $g_{1}$ and $g_{3}$,
in analogy with {\em normal} and {\em Umklapp}
processes in our previous work for
one-dimensional electron system.\cite{ore}
We take only interactions between two electrons with antiparallel spins
into account, as in the Hubbard model, i.e., $g_{1}\equiv g_{1\perp}$ and
$g_{3}\equiv g_{3\perp}$.
We treat the above Hamiltonian 
in the mean field approximation.
We are interested in the repulsive case,
\begin{equation}
0\leq g_{1} \leq g_{3}.\label{repulsive}
\end{equation}

\begin{figure}
\begin{center}
\leavevmode\epsfysize=2cm
\epsfbox{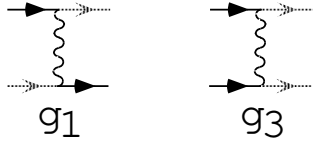}
\end{center}
\caption{The backward scattering processes with large momentum transfer.
The solid and dashed lines stand for electrons in the region 
A and B, 
respectively.}
\label{g1g3}
\end{figure}

We consider three types of the scattering channels:\\
(1) Cooper-pair channel

This channel consists of the scattering processes as shown 
in Fig.~\ref{cchannel} (a), where
a pair of two electrons with total momentum $2Q_{A}$ is scattered to 
that with total momentum  $2Q_{B}$ and vice versa.
This is included only in $g_{3}$ processes. 
We note that this channel is represented 
in the original 1st Brillouin zone as shown in Fig.~\ref{cchannel} (b), 
where there exist 
only {\em normal} processes between two electrons with total
momentum equal to {\em zero}.
\begin{figure}
\begin{center}
\leavevmode\epsfysize=5cm
\epsfbox{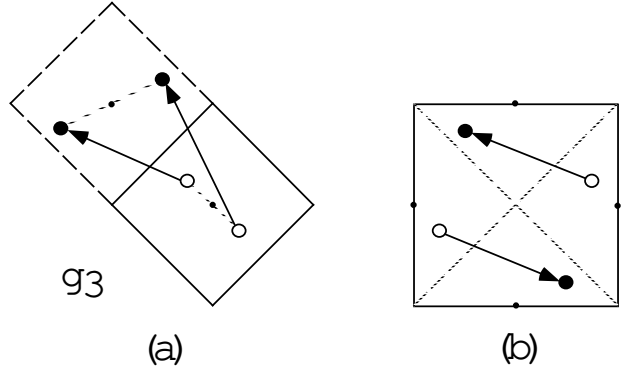}
\end{center}
\caption{The Cooper-pair channel (a) in our Brillouin zone and 
(b) in the original Brillouin zone.}
\label{cchannel}
\end{figure}
Therefore, we naturally introduce the order parameters of the Cooper-pair,
\begin{equation}
\left \{
\begin{array}{ccc}
\Delta_{A}=\mathop{{\sum}'}_{k}<c_{Q_{A}-k\downarrow}
c_{Q_{A}+k\uparrow}>,
\\
\\
\Delta_{B}=\mathop{{\sum}'}_{k}<c_{Q_{B}-k\downarrow}
c_{Q_{B}+k\uparrow}>,
\end{array}
\right .
\end{equation}
\[
\mathop{{\sum}'}_{k} \equiv \sum_{|k_{x}|+|k_{y}|<\pi} 
\equiv \int_{|k_{x}|+|k_{y}|<\pi}\frac{\mbox{d}k_{x}
\mbox{d}k_{y}}{(2\pi)^{2}}.
\]

It is to be noted that $Q_{A}\pm k$ and $Q_{B}\pm k$ always lie in the
region A and B, respectively, for $k$ satisfying $|k_{x}|+|k_{y}|<\pi$
shown in Fig.~\ref{mbz}.
This is the reason for the choice of the Brillouin zone as shown 
in Fig.~\ref{2dbz} (b).

\begin{figure}
\begin{center}
\leavevmode\epsfysize=5cm
\epsfbox{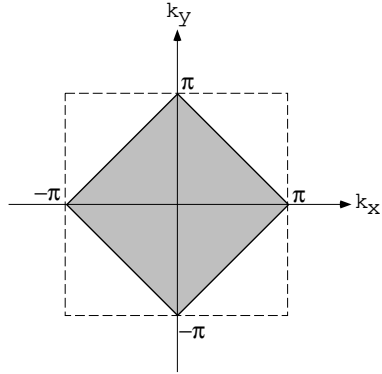}
\end{center}
\caption{The region satisfying $|k_{x}|+|k_{y}|<\pi$ (shaded area).}
\label{mbz}
\end{figure}

For $g_{3}>0$, the order parameters, $\Delta_{A}$ and $\Delta_{B}$,  
can be non-zero only in the case,
\begin{equation}
\Delta_{1}\equiv g_{3}\Delta_{A}=-g_{3}\Delta_{B},\label{gap}
\end{equation}
i.e., the superconducting gap has $d_{x^{2}-y^{2}}$ symmetry:
\begin{eqnarray}
\Delta_{x^{2}-y^{2}}&\equiv&\sum_{p}w_{p}<c_{-p\downarrow}c_{p\uparrow}>,\\
&=&\Delta_{A}-\Delta_{B},\nonumber\\
&=&\frac{2}{g_{3}}\Delta_{1}.\nonumber
\end{eqnarray}

We note that this superconducting gap function is {\em nodeless}
in spite of $d_{x^{2}-y^{2}}$ symmetry, i.e.,
only its sign changes across the boundary between the region A and B
and its magnitude is constant all over the 1st Brillouin zone,
which means that it changes {\em discontinuously} 
along the region boundary.
This can be more clearly seen from the fact that we can rewrite 
(\ref{gap}) as $\Delta(p)=g_{3}\Delta_{1}w_{p}=g_{3}\Delta_{1}
\mbox{sgn} (\cos p_{y}-\cos p_{x})$.
It is due to our choice of the 1st Brillouin zone as Fig.~\ref{2dbz} (b)
and introduction of the 'Umklapp' scattering $g_{3}$ 
with momentum dependence ignored.

This channel is the same as 'pair-tunneling' of two electrons located 
around $Q_{A}$ and $Q_{B}$, which favors $d_{x^{2}-y^{2}}$ pairing,
in the 2D Hubbard model.\cite{kuroki}
These interaction processes are not present as important factors
in the recent model by Assaad {\em et al.}, who
have introduced the additional 
interaction expressed as the square of the
single-particle nearest-neighbor hopping.\cite{imada}\\
(2) density-wave channel

This channel, consisting of the scattering processes 
with momentum transfer equal to $Q\equiv(\pi,\pi)$,
as shown in Fig.~\ref{dchannel} (a), is included
in both $g_{1}$ and $g_{3}$ processes.
In the original 1st Brillouin zone, this is represented
as shown in  Fig.~\ref{dchannel} (b), where there exist
both {\em normal} and {\em Umklapp} processes.
\begin{fullfigure}
\begin{center}
\leavevmode\epsfysize=5cm
\epsfbox{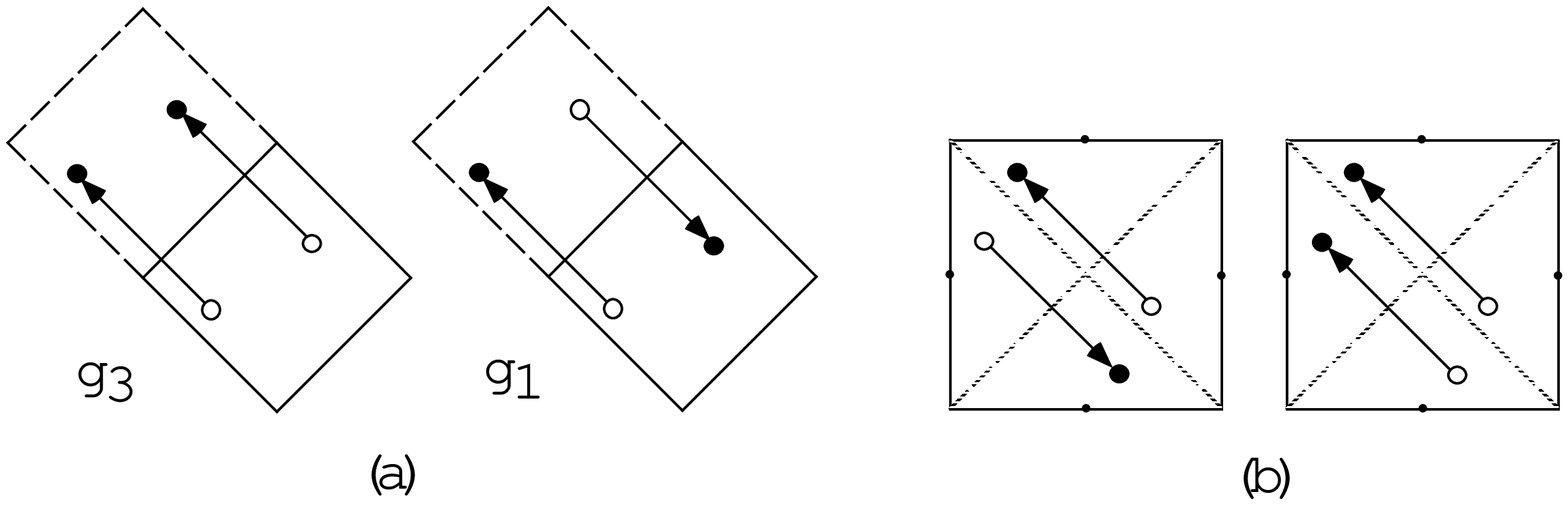}
\end{center}
\caption{The density-wave channel (a) in our Brillouin zone and 
(b) in the original Brillouin zone.}
\label{dchannel}
\end{fullfigure}
We introduce staggered carrier density of each spin,\
\begin{equation}
\left \{
\begin{array}{ccc}
\Delta_{\uparrow}=\mathop{{\sum}'}_{k}<c_{Q_{B}+k\uparrow}^{\dagger}
c_{Q_{A}+k\uparrow}>,\\
\\
\Delta_{\downarrow}=\mathop{{\sum}'}_{k}<c_{Q_{B}-k\downarrow}
^{\dagger}
c_{Q_{A}-k\downarrow}>.
\end{array}
\right .
\end{equation}
For (\ref{repulsive}),
they can be non-zero only in the case,
\[
\Delta_{\uparrow}=-\Delta^{\ast}_{\downarrow}\equiv real,
\]
\begin{equation}
\Delta_{2}\equiv g_{s}\Delta_{\uparrow}=-g_{s}\Delta_{\downarrow},
\end{equation}
where $g_{s}\equiv g_{1}+g_{3}$.
This is the commensurate spin-density-wave (C-SDW) 
or antiferromagnetic (AF) state.
The staggered magnetization (N\'eel order) $M$ is given by
\begin{eqnarray}
M&\equiv&
\frac{1}{N}\sum_{i}(-1)^{i}<S_{i}^{z}>,\\
&=& \frac{1}{2}\sum_{p}<c^{\dagger}_{p+Q\uparrow}c_{p\uparrow}-
c^{\dagger}_{p+Q\downarrow}
c_{p\downarrow}>,\nonumber\\
&=&\mbox{Re}(\Delta_{\uparrow}-\Delta_{\downarrow}),\nonumber\\
&=&\frac{2}{g_{s}}\Delta_{2}.\nonumber
\end{eqnarray}

In this paper, we consider the commensurate case only.
The validity of this assumption will be discussed in \S 4.\\
(3) $\pi$-pair channel

This channel, consisting of the scattering processes between
two electrons with total momentum equal to $Q$, as shown 
in Fig.~\ref{pchannel} (a), is included only in $g_{1}$ processes.
In the original 1st Brillouin zone, this is represented
as shown in  Fig.~\ref{pchannel} (b), where there exist
both {\em normal} and {\em Umklapp} processes 
(only the latter case is shown in the figure).
\begin{figure}
\begin{center}
\leavevmode\epsfysize=5cm
\epsfbox{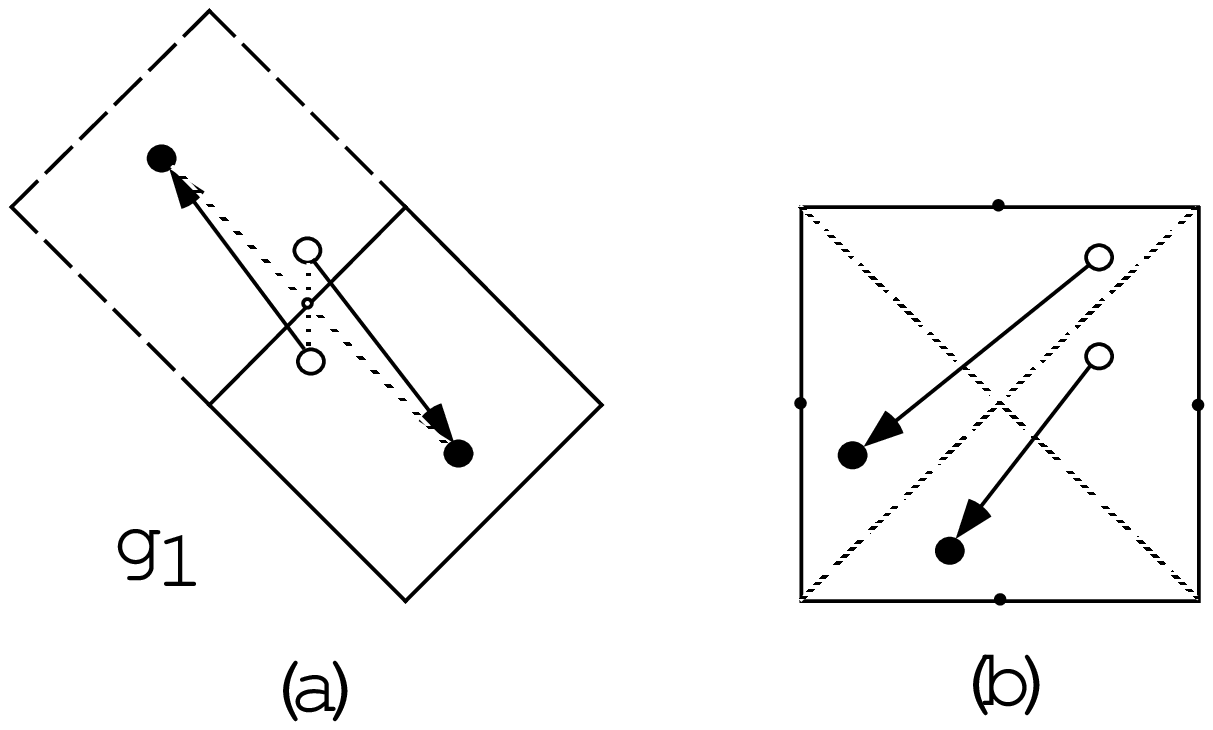}
\end{center}
\caption{The $\pi$-pair channel (a) in our Brillouin zone and 
(b) in the original Brillouin zone.}
\label{pchannel}
\end{figure}
We introduce the order parameters of $\pi$-pair,
\begin{equation}
\left \{
\begin{array}{ccc}
\Delta_{+}=\mathop{{\sum}'}_{k}<c_{Q_{A}-k\downarrow}
c_{Q_{B}+k\uparrow}>,\\
\\
\Delta_{-}=\mathop{{\sum}'}_{k}<c_{Q_{B}-k\downarrow}
c_{Q_{A}+k\uparrow}>.
\end{array}
\right .
\end{equation}
For $g_{1}>0$, they can be non-zero only in the case
\begin{equation}
\Delta_{3}\equiv g_{1}\Delta_{+}=-g_{1}\Delta_{-},
\end{equation}
i.e., $\pi$-{\em triplet} pair exists, 
\begin{eqnarray}
\pi_{triplet}&\equiv&\sum_{p\sigma}w_{p}<c_{-p+Q\sigma}c_{p\bar{\sigma}}>,\\
&=&-2(\Delta_{+}-\Delta_{-}),\nonumber\\
&=&-\frac{4}{g_{1}}\Delta_{3}.\nonumber
\end{eqnarray}
It is necessary to decouple the Hamiltonian by use of 
this order parameter because
if dSC and AF coexist $\pi$-triplet pair 
always results. 
This important fact has been indicated by 
Psaltakis {\em et al.} 
in slightly different context.\cite{crete}
This $\pi$-triplet pair was not taken into account
in the calculation based on the slave-boson mean field approximation 
in the 2D $t$-$J$ model,\cite{inaba} 
where only dSC and AF are simultaneously taken into account.

By use of the above three order parameters,
we obtain the mean field Hamiltonian as follows:
\begin{subequations}
\begin{equation}
H^{MF}=\mathop{{\sum}'}_{k}
\psi^{\dagger}_{k}M_{k}\psi_{k}+E_{c},
\end{equation}
\begin{equation}
\psi^{\dagger}_{k}=(c^{\dagger}_{Q_{A}+k\uparrow},\:
c_{Q_{A}-k\downarrow},\:
c^{\dagger}_{Q_{B}+k\uparrow},\:c_{Q_{B}-k\downarrow}),
\end{equation}
\begin{equation}
M_{k} = \left(
\begin{array}{cccc}
a_{k} & -\Delta_{1} & -\Delta_{2} & \Delta_{3}\\
-\Delta^{\ast}_{1} & -a_{k} & -\Delta^{\ast}_{3} & -\Delta_{2}\\
-\Delta_{2} & -\Delta_{3} & b_{k} & \Delta_{1}\\
\Delta^{\ast}_{3} & -\Delta_{2} & \Delta^{\ast}_{1}  & -b_{k}
\end{array}
\right),
\end{equation}
\begin{equation}
a_{k}\equiv \xi_{Q_{A}\pm k},b_{k}\equiv \xi_{Q_{B}\pm k},
\end{equation}
\begin{equation}
E_{c}=
2\left\{\frac{|\Delta_{1}|^{2}}{g_{3}}+\frac{\Delta_{2}^{2}}{g_{s}}+
\frac{|\Delta_{3}|^{2}}{g_{1}}\right\}-\mu
\end{equation}
\end{subequations}
There are four quasiparticle energy bands, $\pm E_{+}(k)$ and $\pm E_{-}(k)$,
\begin{subequations}
\begin{equation}	
E_{\pm}(k)=\sqrt{\frac{a_{k}^{2}+b_{k}^{2}}{2}+
|\Delta_{1}|^{2}+(\Delta_{2})^{2}+|\Delta_{3}|^{2}
\pm A(k)},
\end{equation}
\begin{full}
\begin{equation}	
A(k)\equiv\sqrt{(a_{k}-b_{k})^{2}\left [ \left ( \frac{a_{k}+b_{k}}{2} \right 
)^{2}+
|\Delta_{3}|^{2} \right ]+[(a_{k}+b_{k})(\Delta_{2})-2\mbox{Re}(\Delta_{1}^{*}
\Delta_{3})]^{2}},
\end{equation}
\end{full}
\end{subequations}
where we note that 
$a_{k}\leftrightarrow b_{k}$, $A(k)\leftrightarrow A(k)$ and
$E_{\pm}(k)\leftrightarrow E_{\pm}(k)$ as we exchange 
$k_{x}$ and $k_{y}$.

The self-consistent equations are given by
\begin{full}
\begin{subeqnarray}
\Delta_{1}&=&
\frac{g_{3}}{4}\mathop{{\sum}'}_{k}\sum_{\alpha = \pm}
\frac{1}{E_{\alpha}(k)}\tanh\frac{E_{\alpha}(k)}{2T} \left \{
\Delta_{1}+\alpha\frac{2\Delta_{3}}{A(k)}\left[\mbox{Re}(\Delta_{1}^{*}
\Delta_{3})-a_{k}\Delta_{2} \right ] \right \},\makebox[3em]{}
\\
\Delta_{2}&=&\frac{g_{s}}{4}\mathop{{\sum}'}_{k}\sum_{\alpha = \pm}
\frac{1}{E_{\alpha}(k)}\tanh\frac{E_{\alpha}(k)}{2T} \left\{
\Delta_{2}\left [1-\alpha\frac{(a_{k}-b_{k})^{2}}{2A(k)}\right ]
-\alpha\frac{2a_{k}}{A(k)}\left[\mbox{Re}(\Delta_{1}^{*}
\Delta_{3})-a_{k}\Delta_{2} \right ]\right\},\makebox[3em]{}
\\
\Delta_{3}&=&\frac{g_{1}}{4}\mathop{{\sum}'}_{k}\sum_{\alpha = \pm}
\frac{1}{E_{\alpha}(k)}\tanh\frac{E_{\alpha}(k)}{2T} \left\{
\Delta_{3}\left [1+\alpha\frac{(a_{k}-b_{k})^{2}}{2A(k)}\right ]
+\alpha\frac{2\Delta_{1}}{A(k)}\left[\mbox{Re}(\Delta_{1}^{*}
\Delta_{3})-a_{k}\Delta_{2} \right ]\right\},\makebox[3em]{}
\\
n&=&1-\mathop{{\sum}'}_{k}\sum_{\alpha = \pm}
\frac{1}{E_{\alpha}(k)}\tanh\frac{E_{\alpha}(k)}{2T} \left \{
a_{k}\left [ 1+\alpha\frac{(a_{k}-b_{k})^{2}}{2A(k)}\right ]
-\alpha\frac{2\Delta_{2}}{A(k)}\left [ \mbox{Re}(\Delta_{1}^{*}
\Delta_{3})-a_{k}\Delta_{2}\right ] \right\},\makebox[3em]{}
\label{sce}
\end{subeqnarray}
\end{full}
where $n$ is the electron filling, which is related to the hole doping $\delta$
and the electron doping $x$
by $\delta = 1-n$ and $x=n-1$, respectively.
These equations satisfy $\frac{\partial F}{\partial \Delta_{1}}
=\frac{\partial F}{\partial \Delta_{2}}
=\frac{\partial F}{\partial \Delta_{3}}
=0$, where
$F$ is the free energy per one lattice site in the canonical ensemble
given by
\begin{eqnarray}
F&=&-2T\mathop{{\sum}'}_{k}\left \{ \log\left [ 2\cosh\frac{E_{+}(k)}{2T}
\right ]
+\log\left[2\cosh\frac{E_{-}(k)}{2T}\right ]
\right \}\nonumber\\
&&+E_{c}+\mu n.\label{fene}
\end{eqnarray}

The electronic density of states $\rho (\epsilon)$ is given by
\begin{eqnarray}
\rho(\epsilon)&=&\mathop{{\sum}'}_{k}\sum_{\alpha, \alpha^{'} = \pm}
\delta(\epsilon-\alpha^{'}E_{\alpha}(k))\nonumber\\
&\times&\left \{  1+\frac{\alpha^{'}}
{E_{\alpha}(k)}\left \{ a_{k}\left [ 1+\alpha\frac{(a_{k}-b_{k})^{2}}{2A(k)}
\right ] \right .\right .\nonumber\\
&-& \left .\alpha
\frac{2\Delta_{2}}{A(k)}\left [\mbox{Re}(\Delta_{1}^{*}
\Delta_{3})-a_{k}\Delta_{2}\right ] \right \} \Biggr\},\label{dos}
\end{eqnarray}
which is related with $n$ 
by $n=\int_{-\infty}^{\infty}$d$\epsilon
\rho (\epsilon)
f(\epsilon,T)$, where $f(\epsilon,T)=(\mbox{e}^{\epsilon / T}+1)^{-1}$
is the Fermi distribution function. 

\section{Results}
The self-consistent equations (\ref{sce}) are solved numerically
and the phase diagram has been determined
in the plane of temperature $T$ and the doping rate $\delta$ or $x$.
It is easily understood from (\ref{sce}) that 
[1] if two of the three order parameters coexist,
another one always results and  
[2] for $g_{1}/t=0$, $\pi$-triplet pair cannot exist
($\Delta_{3}\equiv 0$), i.e., there exists no coexistent state and 
dSC prevails over AF.
We note that it is energetically favorable to take 
both $\Delta_{1}$ and $\Delta_{3}$ to be real for the coexistent state.
We choose $t'$ and $t''$ so that the bare band width $W$ is equal to $8t$
and reproduce various shapes of the Fermi surface. 
We fix the value $g_{3}/t=5.0$.

In the numerical calculation of the DOS (\ref{dos}),
we approximate $\delta$-function by
\begin{equation}
\delta(\epsilon)=-\lim_{\eta\rightarrow0+}\frac{\partial}{\partial \eta}
f(\epsilon,\eta)=\lim_{\eta\rightarrow0+}\{ 4\eta\cosh ^{2}\frac{\epsilon}
{2\eta}\}
^{-1},
\end{equation}
with a {\em finite} value $\eta/t\equiv0.01$.

First we include $t$ only ($t'/t=t''/t=0$).
In this case, there is perfect nesting property 
$\epsilon_{p}=-\epsilon_{p+Q}$ or $b_{k}=-a_{k}-2\mu$.
Since the quasiparticle dispersion depends on $k$ 
only through $a_{k}$,
it is convenient in the numerical calculation
to rewrite (\ref{sce}), (\ref{fene}) and (\ref{dos})
by use of the following identity for any function $g$ of $a_{k}$,
\begin{equation}
\mathop{{\sum}'}_{k}g(a_{k})\equiv\frac{1}{2}\int\mbox{d}\epsilon
\rho_{0}(\epsilon)g(\epsilon-\mu).
\end{equation}
$\rho_{0}(\epsilon)$ is the bare DOS given by
\begin{eqnarray}
\rho_{0}(\epsilon)&\equiv&\sum_{p}\delta (\epsilon-\xi_{p}),
\nonumber\\
&=& \frac{1}{2\pi^{2}t}K\left (\sqrt{1-(\frac{\epsilon}{4t})^{2}}
\right )
\theta(4t-|\epsilon|),
\end{eqnarray}
where $K$ is the complete elliptic integral 
of the 1st kind $K(k)\equiv \int_{0}^{1}
\frac{{\rm d}x}{\sqrt(1-x^{2})(1-k^{2}x^{2})}$ and $\theta$ is the
step function.

We show the bare Fermi surface in Fig.~\ref{fsnesting}
for a few values of the hole doping $\delta$ (it is sufficient to consider the 
hole-doped case due to the particle-hole symmetry).
The saddle points lie on the Fermi surface in the half-filled case.
\begin{figure}
\begin{center}
\leavevmode\epsfysize=5cm
\epsfbox{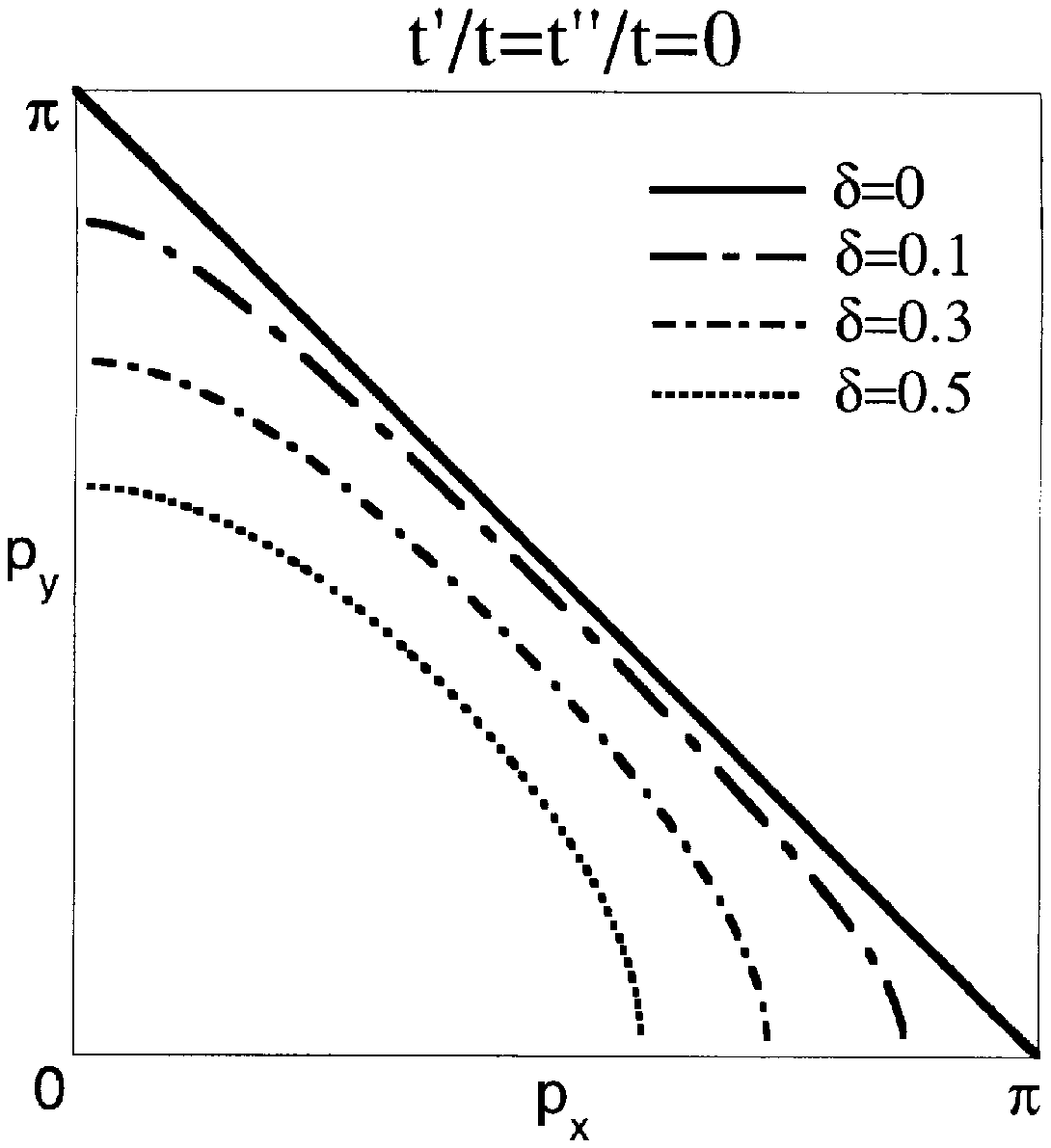}
\end{center}
\caption{The Fermi surface in the case without $t'$ and $t''$
for a few values of the hole doping $\delta$.}
\label{fsnesting}
\end{figure}

We show the phase diagram in Fig.~\ref{pdnt} for
two different choices of the coupling constants,
$g_{1}=g_{3}$ and $g_{1}=g_{3}/2$.
In the half-filled case, as the temperature is lowered,
AF gets stabilized, and since there is no free carriers 
in the presence of the SDW gap,
dSC cannot arise even in $T=0$.
On the other hand, near half filling, there are free carriers
even in the AF state, and dSC also can be stabilized in the low temperature
region. This result is similar to the spin-bag picture.\cite{spinbag}
It is important that the coexistence of AF and dSC always results 
in $\pi$-triplet pair.
Such a close relationship between dSC and AF has been indicated
by SO(5) theory.\cite{SCZhang}
For smaller $g_{1}$, AF region and therefore the coexistent region
are reduced.

We note that for $g_{1}=0$ and $\delta=0$, 
due to the perfect nesting property,
$\Delta_{1}$ and $\Delta_{2}$ are determined by the same equation,
i.e., AF and dSC are degenerate.

\begin{fullfigure}
\begin{center}
\leavevmode\epsfysize=6cm
\epsfbox{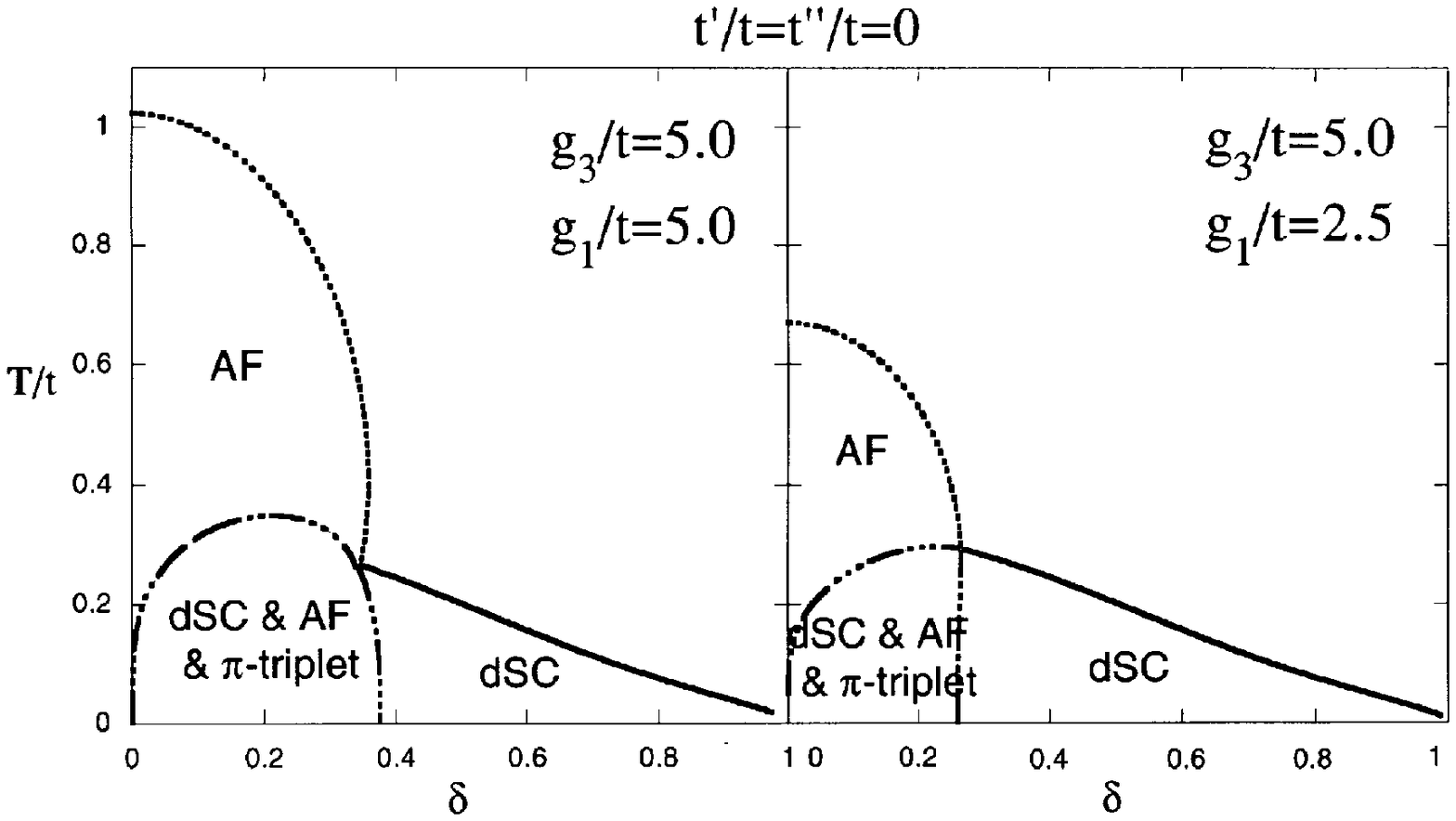}
\end{center}
\caption{The phase diagram in the case without $t'$ and $t''$
for $g_{3}/t=5.0$ and $g_{1}/t=5.0,2.5$.}
\label{pdnt}
\end{fullfigure}

In Fig.~\ref{nesdos}
we show the DOS in the coexistent state near the optimal doping which gives
the highest onset temperature of dSC.
There exists energy gap in the DOS at the Fermi level.

\begin{figure}
\begin{center}
\leavevmode\epsfysize=5cm
\epsfbox{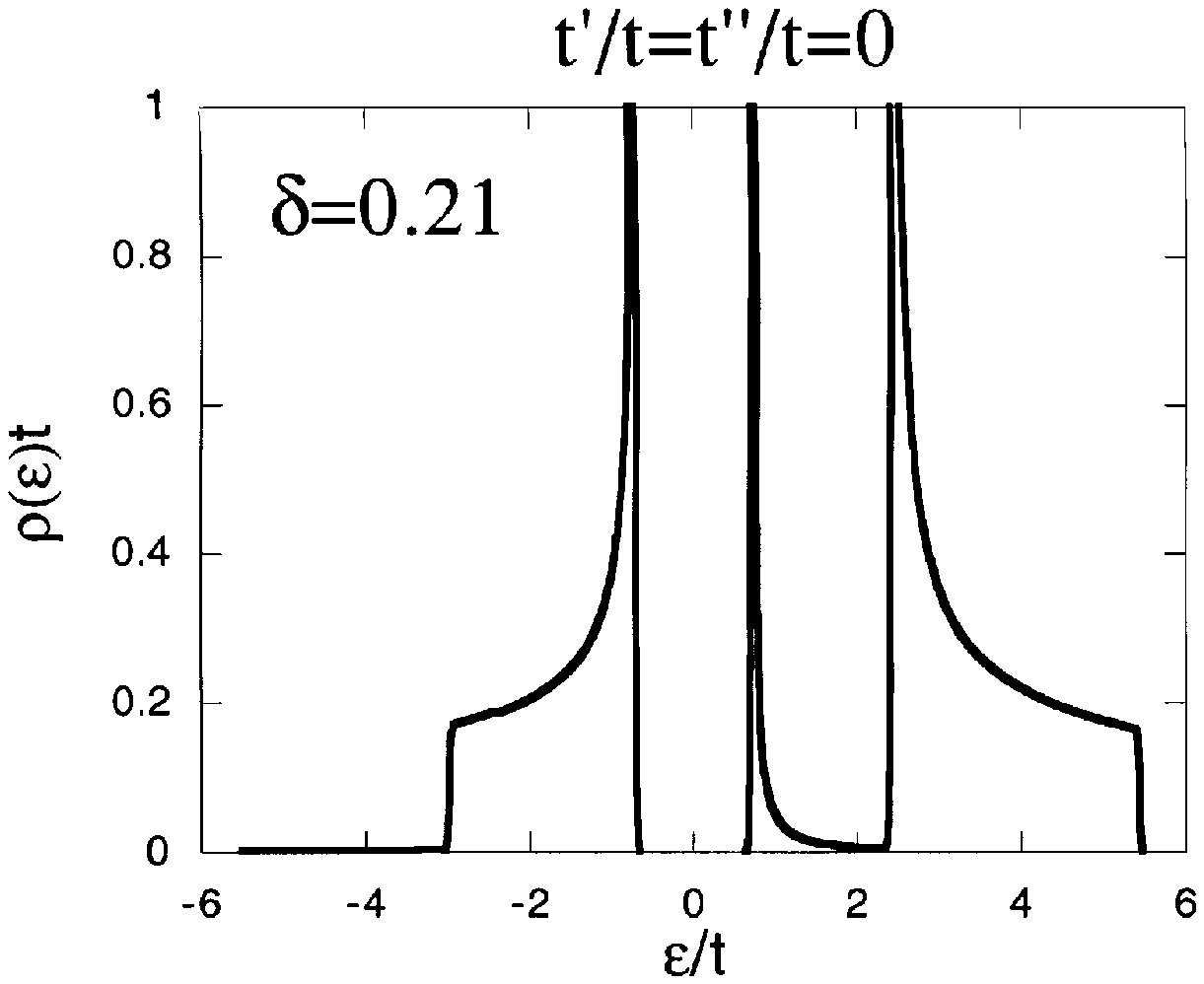}
\end{center}
\caption{The electronic DOS in the case without $t'$ and $t''$
in the coexistent state 
for $g_{3}/t=g_{1}/t=5.0$ and $T=0$ near the optimal doping $\delta=0.21$.
The Fermi level lies at $\epsilon=0$.}
\label{nesdos}
\end{figure}

Next we consider the hole-doped case with the YBCO type Fermi surface.
We choose $t'/t=-1/5$ and $t''/t=1/7$.
The Fermi surface is shown in Fig.~\ref{fsYBCO}
for a few values of the hole doping $\delta$.
In this case, the saddle points lie on the Fermi surface 
near $\delta=0.44$.
\begin{figure}
\begin{center}
\leavevmode\epsfysize=5cm
\epsfbox{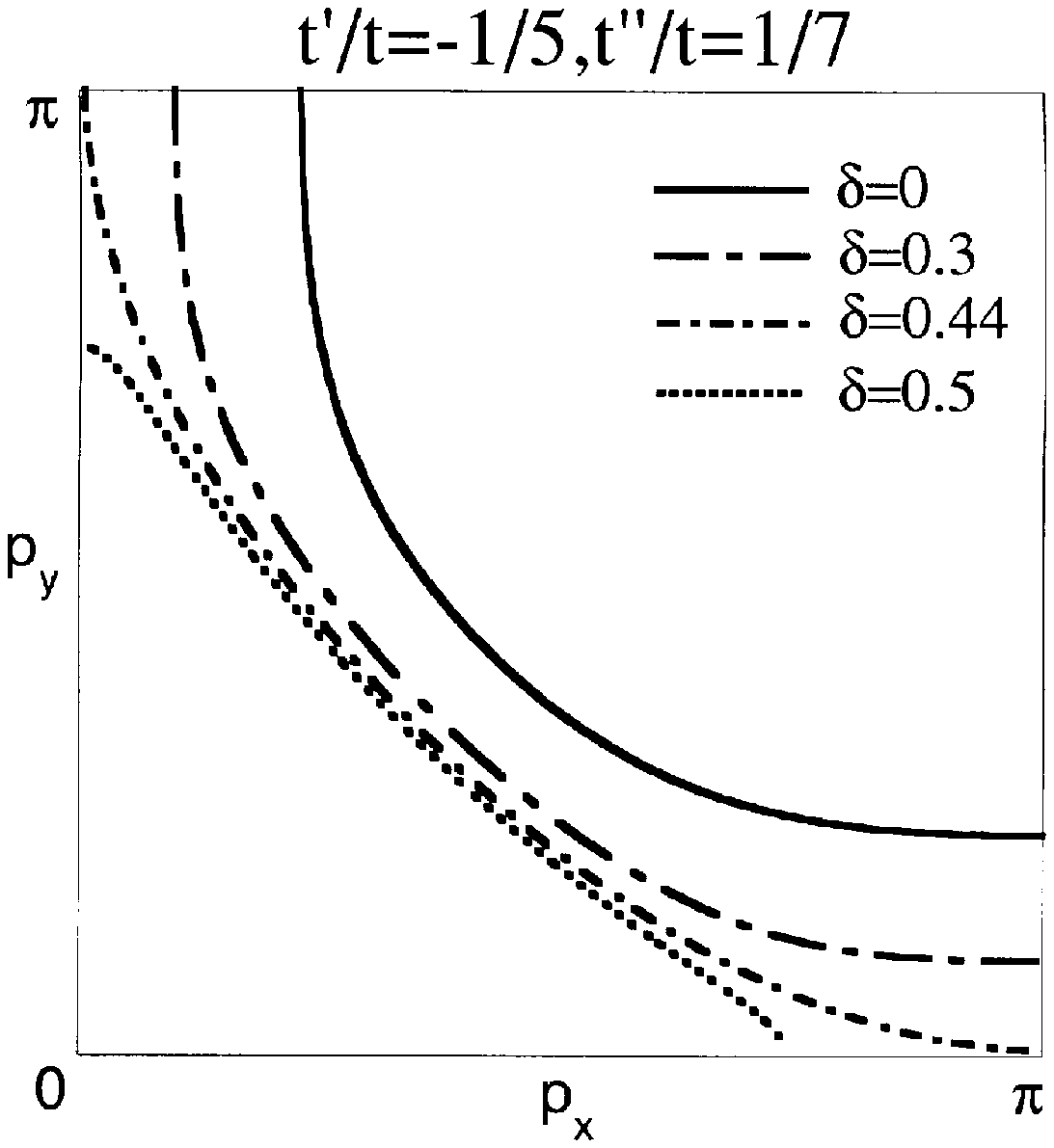}
\end{center}
\caption{The Fermi surface for the YBCO case
for a few values of the hole doping $\delta$.}
\label{fsYBCO}
\end{figure}

We show the phase diagram in Fig.~\ref{pdh1} for
two different choices of the coupling constant,
$g_{1}=g_{3}$ and $g_{1}=g_{3}/2$.
For $\delta=0$, only AF can be stabilized, and
near half filling there exists coexistent state of
dSC, AF and $\pi$-triplet pair. 
For $g_{1}=g_{3}$,
compared with the perfect nesting case, 
the effect of nesting near $\delta=0$ is suppressed and the onset temperature
of AF is lower. On the other hand, near $\delta\sim 0.4$,
the saddle points lying near the Fermi surface lead to the enhancement
of the order parameters of both dSC and AF.
\begin{fullfigure}
\begin{center}
\leavevmode\epsfysize=6cm
\epsfbox{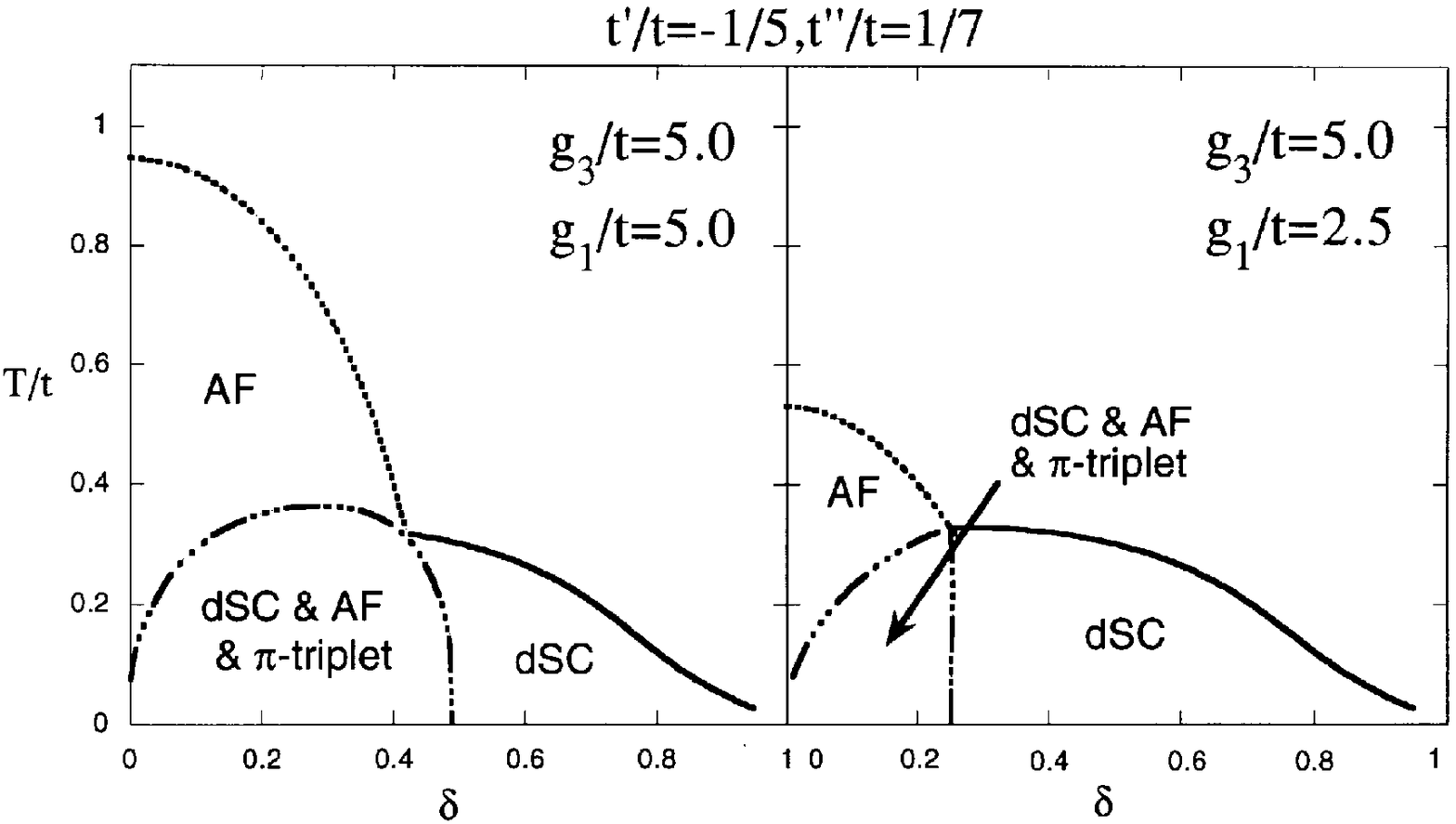}
\end{center}
\caption{The phase diagram for the YBCO case
for $g_{3}/t=5.0$ and $g_{1}/t=5.0,2.5$.}
\label{pdh1}
\end{fullfigure}

In Fig.~\ref{h1dos}
we show the DOS in the coexistent state near the optimal doping.
The DOS has an energy gap at the Fermi level. 
\begin{figure}
\begin{center}
\leavevmode\epsfysize=5cm
\epsfbox{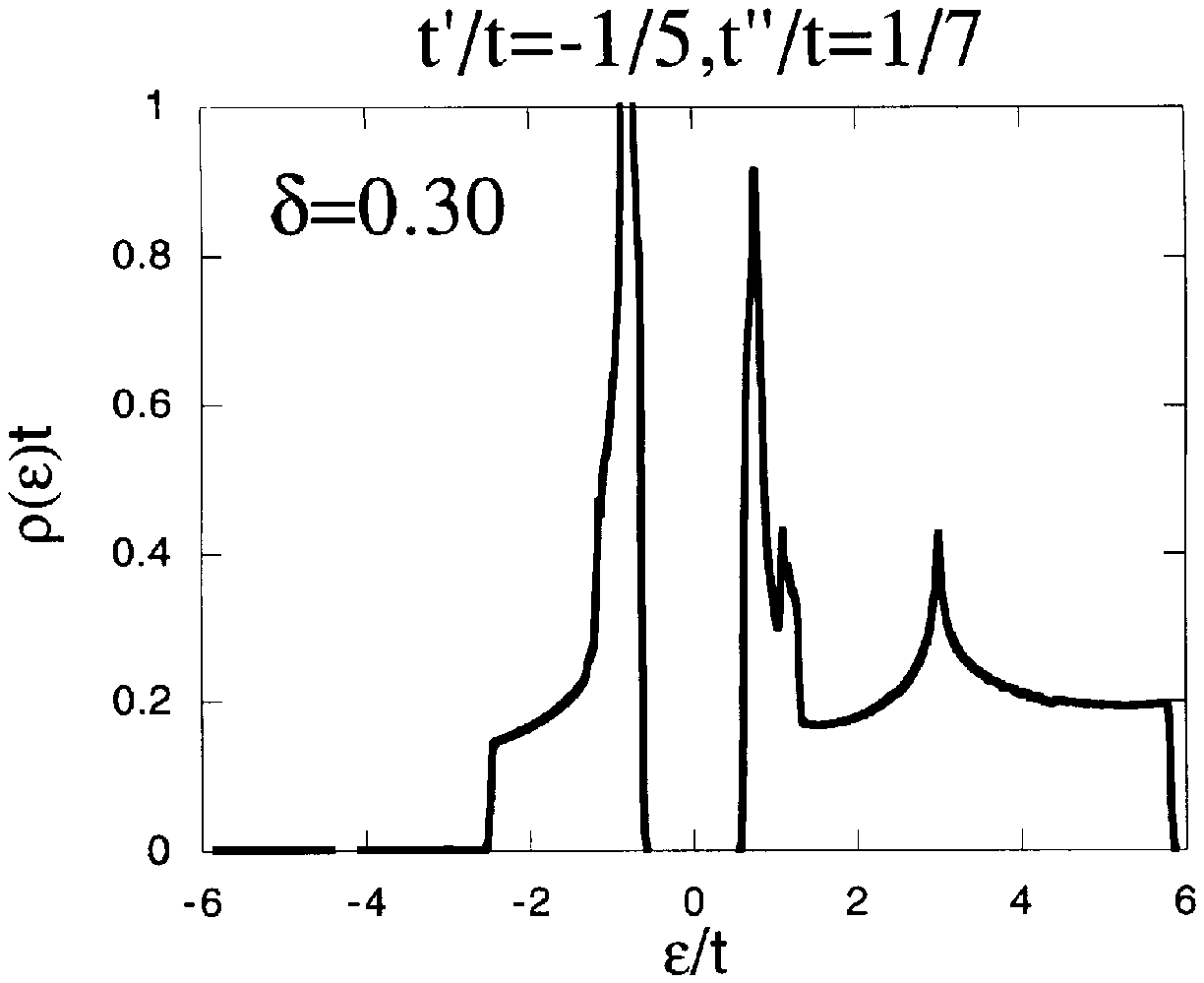}
\end{center}
\caption{The electronic DOS for the YBCO case
in the coexistent state 
for $g_{3}/t=g_{1}/t=5.0$ and $T=0$ near the optimal doping $\delta=0.30$.
The Fermi level lies at $\epsilon=0$.}
\label{h1dos}
\end{figure}

Finally we consider the electron-doped case with the NCCO type Fermi surface.
We choose $t'/t=-1/3$ and $t''/t=1/10$.
The Fermi surface is shown in Fig.~\ref{fsNCCO}
for a few values of the electron doping $x$.
In this case, the saddle points lie on the Fermi surface 
near $x=0.28$.
\begin{figure}
\begin{center}
\leavevmode\epsfysize=5cm
\epsfbox{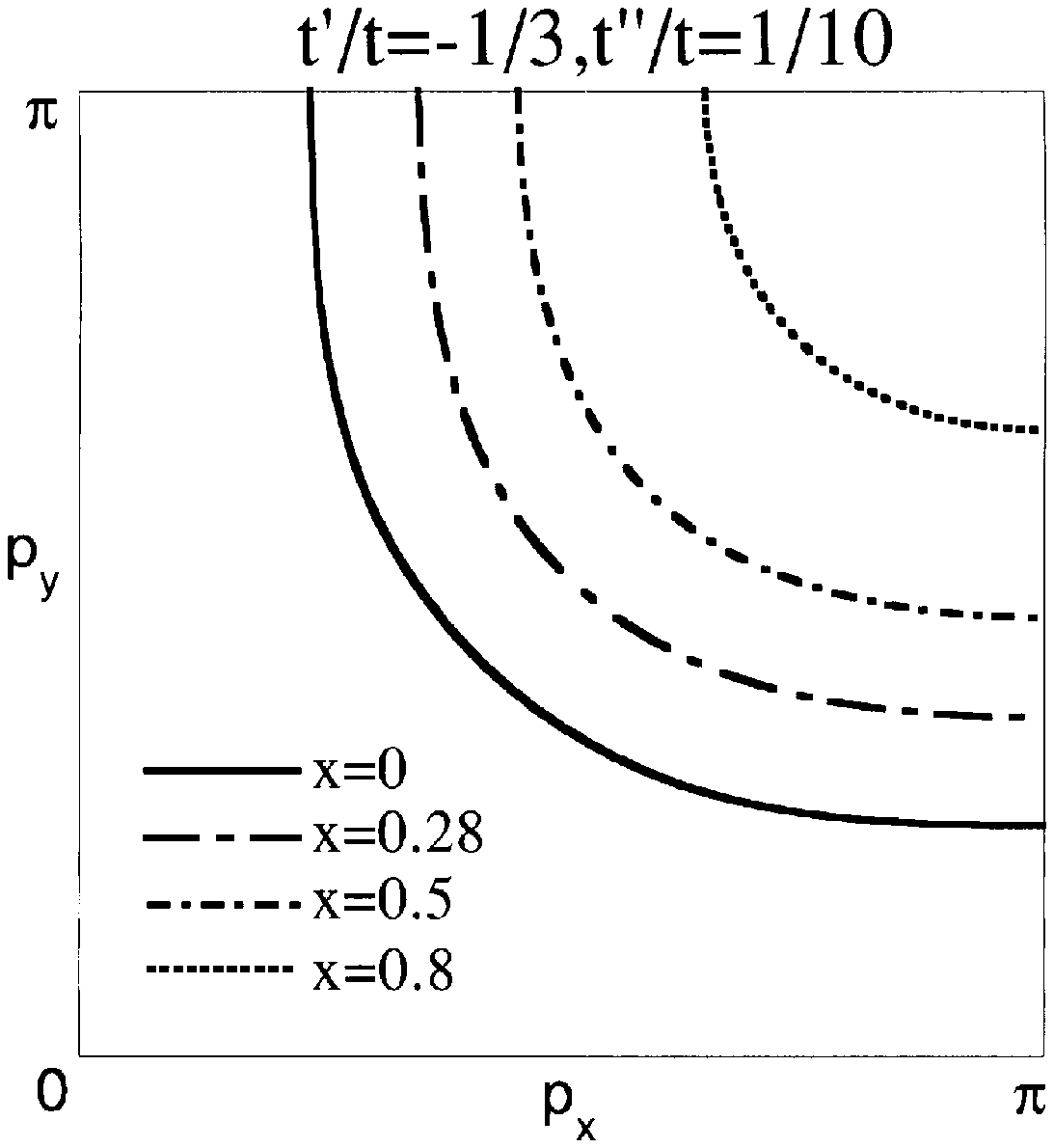}
\end{center}
\caption{The Fermi surface for the NCCO case
for a few values of the electron doping $x$.}
\label{fsNCCO}
\end{figure}

We show the phase diagram in Fig.~\ref{pdel} for
two different choices of the coupling constant,
$g_{1}=g_{3}$ and $g_{1}=g_{3}/2$.
In this case the saddle points are not located near the Fermi surface
and there is a suppression of the onset temperature of AF and dSC
compared with the above two cases.
\begin{fullfigure}
\begin{center}
\leavevmode\epsfysize=6cm
\epsfbox{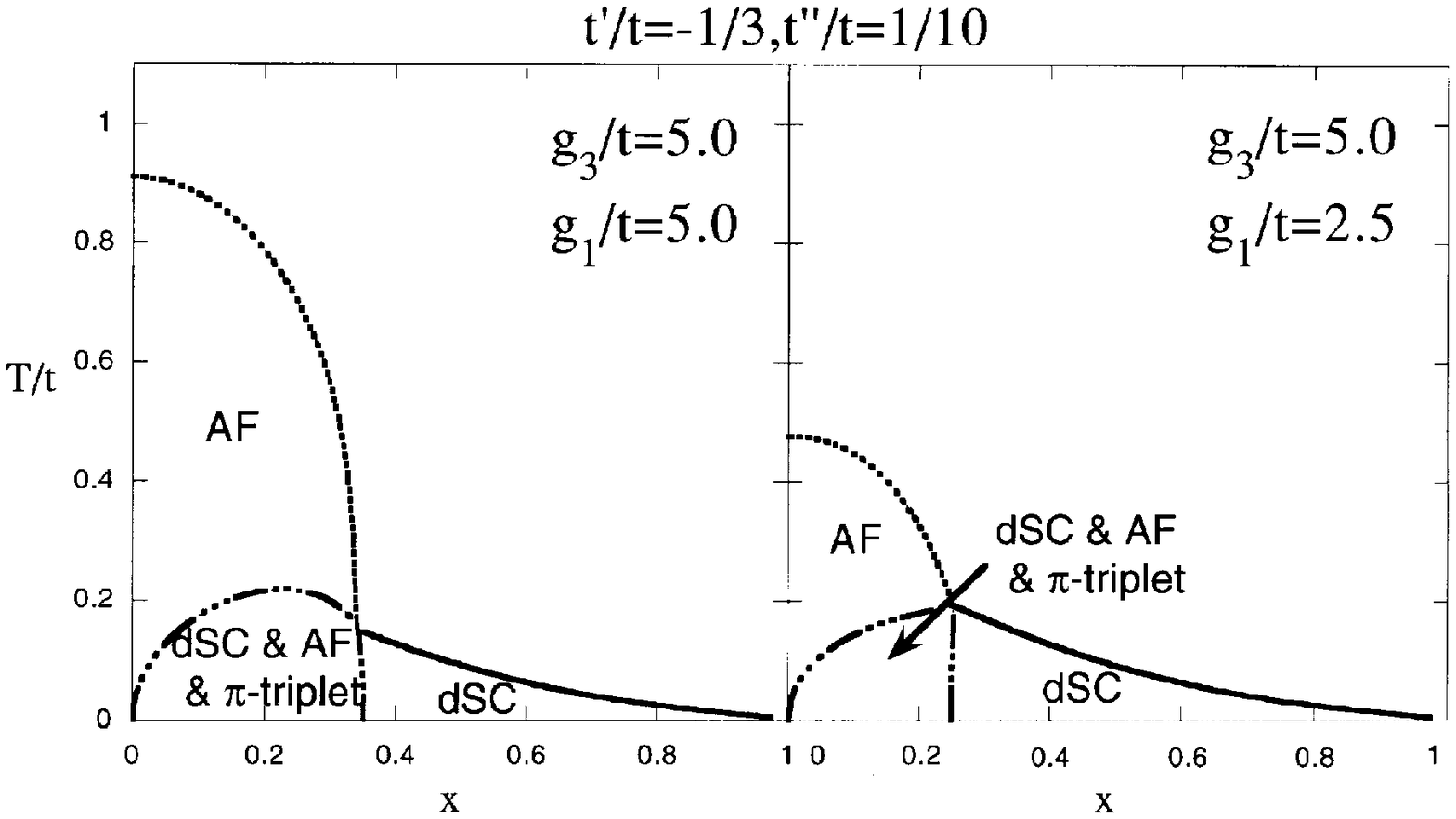}
\end{center}
\caption{The phase diagram for the NCCO case
for $g_{3}/t=5.0$ and $g_{1}/t=5.0,2.5$.}
\label{pdel}
\end{fullfigure}

In Fig.~\ref{eldos}
we show the DOS in the coexisting state near the optimal doping.
The energy gap lies at the Fermi level in the DOS.
\begin{figure}
\begin{center}
\leavevmode\epsfysize=5cm
\epsfbox{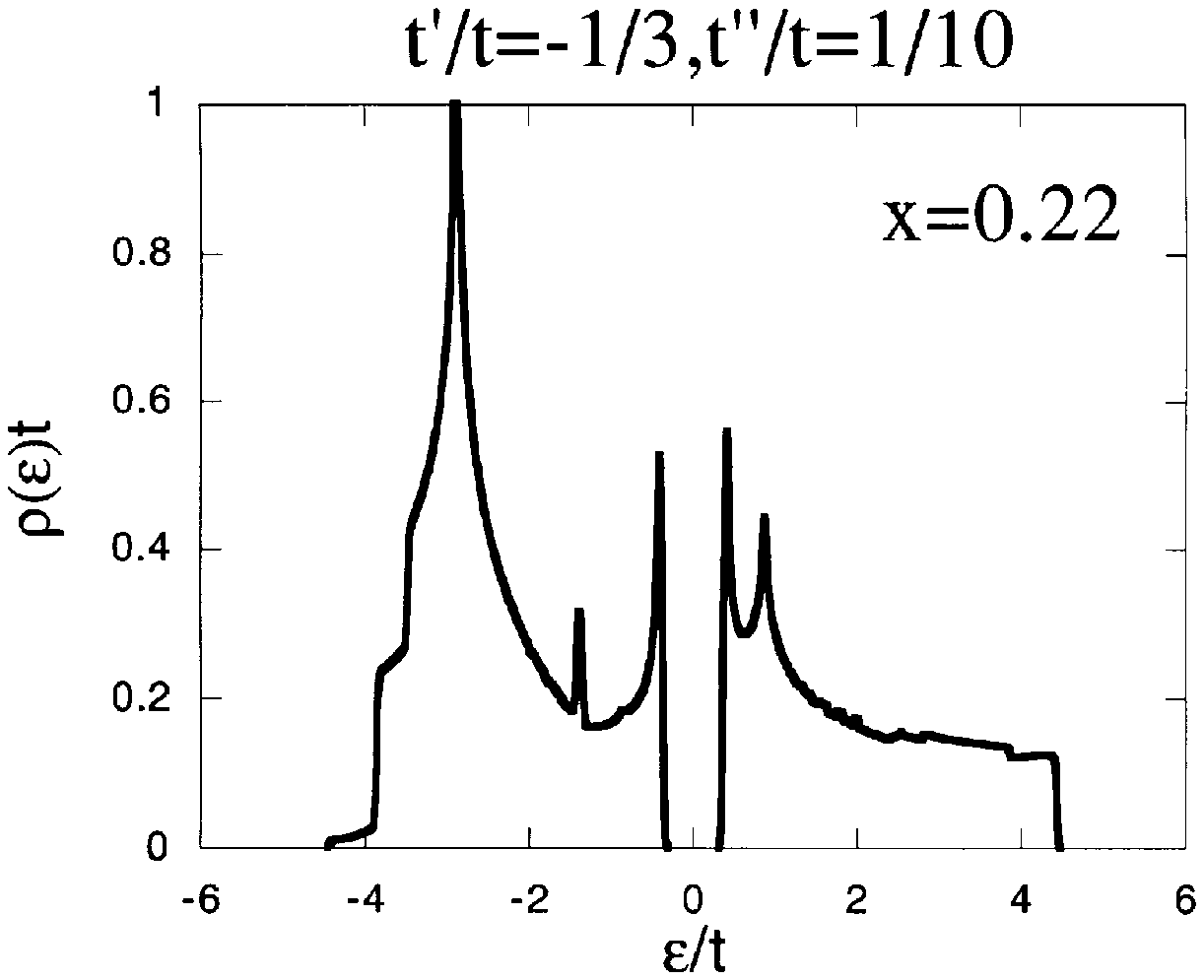}
\end{center}
\caption{The electronic DOS for the NCCO case
in the coexistent state 
for $g_{3}/t=g_{1}/t=5.0$ and $T=0$ near the optimal doping $x=0.22$.
The Fermi level lies at $\epsilon=0$.}
\label{eldos}
\end{figure}

\section{Conclusion}
We have studied possible ordered states 
of interacting electrons on a square lattice
with a special emphasis
on the backward scattering, $g_{1}$ and $g_{3}$, i.e.,
'exchange' and 'Umklapp' processes, respectively, between two electrons around
($\pi$,0) and (0,$\pi$) for various shapes of the Fermi surface.

We focus on the case that $g_{3}$, which is related with
the superconducting state with $d_{x^{2}-y^{2}}$ type of the gap, 
is about half of the bare band width, $W/t=8.0$, i.e., $g_{3}/t$=5.0. 
For $g_{1}=0$, $d$-wave superconducting state (dSC)
competes with and prevails over antiferromagnetism (AF)
even in the half-filed case.
For $0<g_{1}\leq g_{3}$ with $g_{1}/g_{3}$ not so small, 
only AF can be stabilized
in the half-filled case and
coexist with dSC near half filling.
Especially in the coexistent state, 
$\pi$-triplet pair always results and 
the DOS has an energy gap at the Fermi level,
independent of the hole or electron doping rate.
The above conclusion is qualitatively independent of the shapes of the
Fermi surface. 

As we have indicated in \S 2, we have assumed C-SDW
throughout this paper.
In general, however, away from half filling, 
the {\em incommensurate} SDW (IC-SDW) should be expected
with $Q\ne(\pi,\pi)$.
Therefore, it is not clear whether this coexistent state 
can be stabilized or not when we include IC-SDW.
In fact, the instability towards incommensurate IC-SDW
depends strongly
on momentum-dependence of both
the zero-frequency susceptibility, $\chi_{0}(q)$, and effective interaction.
We note, however, the following facts:
[1] There exists the region where $\chi_{0}(q)$ shows maximum 
at $q=(\pi,\pi)$ near half filling at low temperature, in the present choices
of the Fermi surface.
[2] $g_{1}$ and $g_{3}$ can be considered to have maximum at
$q=(\pi,\pi)$ ($q$ stands for the momentum transfer of one electron),
although the $q$-dependence of $g_{1}$ and $g_{3}$,
which is not explicitly indicated in this paper,
is irrelevant under the assumption of C-SDW of interest.
These facts result in RPA susceptibility showing maximum at $q=(\pi,\pi)$
leading to C-SDW at least near half filling at low temperature where
the coexistent state is possible.

With respect to the superconducting gap symmetry,
in the case that ($\pm \pi/2$,$\pm \pi/2$) lies near the Fermi surface,
not only $d_{x^{2}-y^{2}}$ component but also
other ones can mix.
Recent QMC calculations show that 
not only $d_{x^{2}-y^{2}}$ but also $d_{xy}$ pairing correlation 
are enhanced in the 2D Hubbard model,\cite{kuroki}
and new ordered state with a mixing of 
extended $s$-wave component with total momentum ($\pi$,0) and (0,$\pi$)
and the usual $d_{x^{2}-y^{2}}$ component
has been discovered
in the 2D $t$-$J$ model, in the mean field and 
Gutzwiller approximations.\cite{ogata}
Their results are closely related to the fact that
the original $d_{x^{2}-y^{2}}$-wave superconducting
gap has nodes around 
($\pm \pi/2$,$\pm \pi/2$), i.e.,
the fact that Umklapp scattering between two quasiparticles 
on the Fermi surface around these points is possible
may lead to form the full gap all over the Fermi surface.
Our $d_{x^{2}-y^{2}}$-wave superconducting gap, however, 
has {\em no} node, i.e., changes discontinuously along the boundary
between the region A and B, 
as already discussed in \S 2.
Therefore, in our model,
we cannot discuss the mixing of another component
in the superconducting gap function
in addition to the $d_{x^{2}-y^{2}}$-wave component.
This is the future problem.

\section*{Acknowledgements}
M.M. would like to express his gratitude to Hiroshi Kohno and Kazuhiko
Kuroki for instructive 
discussions and suggestions.  
This work is financially supported by a Grant-in-Aid
for Scientific Research on Priority Area "Anomalous Metallic State near the 
Mott Transition" (07237102) from the Ministry of Education, Science, Sports
and Culture.

\end{document}